# Design of a Ternary Edge-Triggered D Flip-Flap-Flop for Multiple-Valued Sequential Logic

Reza Faghih Mirzaee [1*], Niloofar Farahani [2]

[1] Department of Computer Engineering, Shahr-e-Qods Branch, Islamic Azad University, Tehran, Iran

[2] ECE Department, K. N. Toosi University of Technology, Tehran, Iran

[*] E-Mail: r.f.mirzaee@qodsiau.ac.ir

**Abstract:** Development of large computerized systems requires both combinational and sequential circuits. Registers and counters are two important examples of sequential circuits, which are widely used in practical applications like CPUs. The basic element of sequential logic is Flip-Flop, which stores an input value and returns two outputs ($Q$ and $\overline{Q}$). This paper presents an innovative ternary D Flip-Flap-Flop, which offers circuit designers to customize their design by eliminating one of the outputs if it is not required. This unique feature of the new design leads to considerable power reduction in comparison with the previously presented structures. The proposed design is simulated and tested by HSPICE and 45 nm CMOS technology.

## 1. Introduction

Boolean algebra has indeed played an important role in building computerized systems. The significance of binary numeral system is mainly because of the two-valued nature of electronic components, which are open or closed, connected or disconnected, switched on or off. However, a binary device can merely represent two states. For example, a single-bit memory cell can only hold two logic values of ‘0’ and ‘1’, but not more. Multiple-Valued Logic (MVL) aims to compact more information in a single gate or wire by going beyond dualism, and respond to VLSI and real-world applications more efficiently [1, 2]. Its advantages have been confirmed in many applications such as memories, communications, and digital signal processing [3-5].

The most efficient radix for computing and implementing of switching systems is $e$ (≈2.71828), which makes 3 the best integral value [6]. Theoretically, ternary computations are performed a factor of $\log_2 3$ (≈1.585) faster than binary. The unbalanced ternary with the number set of $\{0, 1, 2\}_3$ is the most common representation, and it is known to be an extension to binary logic [7]. Ternary digits (Trits) are implemented in digital electronics by the voltage levels 0V, $\frac{1}{2}V_{DD}$, and $V_{DD}$, with the common assumption of having only two supply rails, $V_{DD}$ and GND. Thus, the logic value ‘1’ is the result of a simple voltage division between $V_{DD}$ and GND.

The realization of large ternary systems depends on the availability of ternary codes and algorithms as well as design and implementation of electronic circuits. Several ternary logical gates and arithmetic components have previously been designed [8-11]. The design and realization of different ternary memory cells like Static [12, 13] and Dynamic [14] Random Access Memories (SRAM and DRAM), Content-Addressable Memory (CAM) [15], and Flip-Flap-Flop (FFF) [16, 17] have been reported in the literature as well.

Computer circuits consist of both combinational and sequential components. The availability of sequential circuits is a must for developing any practical system. The most promising applications of MVL are memories and arithmetic circuits [7]. In addition, the most essential element of sequential circuitry is Flip-Flop (FF), which is a single-bit memory cell with two stable states. The other practical sequential circuits such as registers and counters are based on this fundamental building block. In ternary logic, the memory cell must be able to hold three different values. The name Flip-Flap-Flop reflects this fact and represents three states.

A single-trit R-S latch has been presented in [16] by cross-coupling two ternary NAND gates. Then, it is converted to a ternary D Flip-Flap-Flop (Fig. 1a). Another ternary D F.F.F. has been introduced in [17] by using cross-coupled ternary inverters (Fig. 1b). Both designs are in accord with the standard sequential design process similar to what it happens in binary logic. They use ternary gates to form a memory cell. High static power consumption is their major drawback, mainly because every single ternary component individually dissipates considerable power. This paper includes static power analysis for the mentioned designs.

In this paper, an innovative latch is realized by using binary logical gates. It is efficiently convertible into a ternary latch. Then, the entire memory cell is transformed into a ternary edge-triggered D Flip-Flap-Flop. Binary logical gates do not consume as much power as ternary counterparts do. The unbalanced ternary components divide voltage in order to produce logic '1'. It causes considerable power dissipation. The correct functionality is tested by simulating the new design with 45 nm Metal-Oxide-Semiconductor Field Effect Transistors (MOSFETs).

The rest of the paper is organized as follows: The new ternary edge-triggered D Flip-Flap-Flop is proposed in Section 2 step by step through four distinctive subsections. Simulation results and comparisons are given in Section 3. Finally, Section 4 concludes the paper.

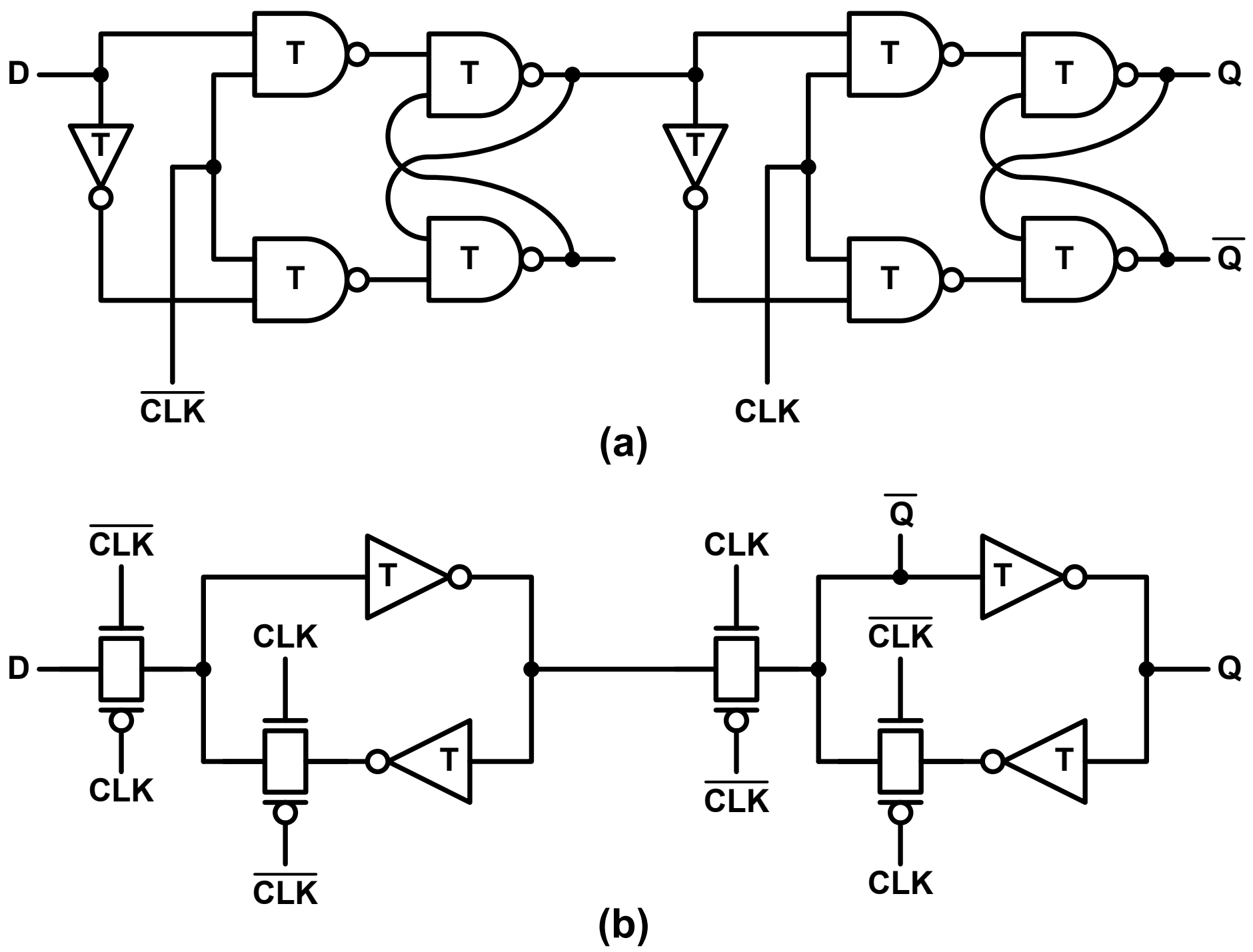

Fig. 1. The previous ternary D Flip-Flap-Flops, (a) Presented in [16], (b) Presented in [17]

## 2. The Proposed Design

The basic memory core in sequential logic is called *latch*. The well-known R-S latch is a pair of cross-coupled NOR gates. There are three other types of latch, known as J-K, T, and D. The latter stores an input value and is widely used in sequential circuits. Flip-Flops are built by adding a clock signal (CLK) to the latches so that the operation is synchronized by either the level (level-sensitive) or the edge (edge-sensitive) of the clock signal.

*2.1 The Proposed Ternary Latch*

The idea of the proposed design is based on a binary latch composed of one AND (Eq. 1) and one OR (Eq. 2) as it is illustrated in Fig. 2a. The inner structure of the conventional Flip-Flops is built in the way that two complementary outputs are simultaneously generated. Unlike them, the binary output values of the proposed design ($Q_1$ and $Q_2$) do not comply with this norm. They sometimes have equal values. In fact, this is the key feature of the proposed design, which enables us to convert it to a ternary latch. Binary values are indicated by either *'Low'* and *'High'* or 0 and $V_{DD}$ in this paper in order not to mix them up with ternary digits '0', '1', and '2'. The block diagram of the proposed ternary latch is depicted in Fig. 2b.

$$Q_1(t) = S_1 . Q_2(t - \Delta t) \tag{1}$$

$$Q_2(t) = S_2 + Q_1(t - \Delta t) \tag{2}$$

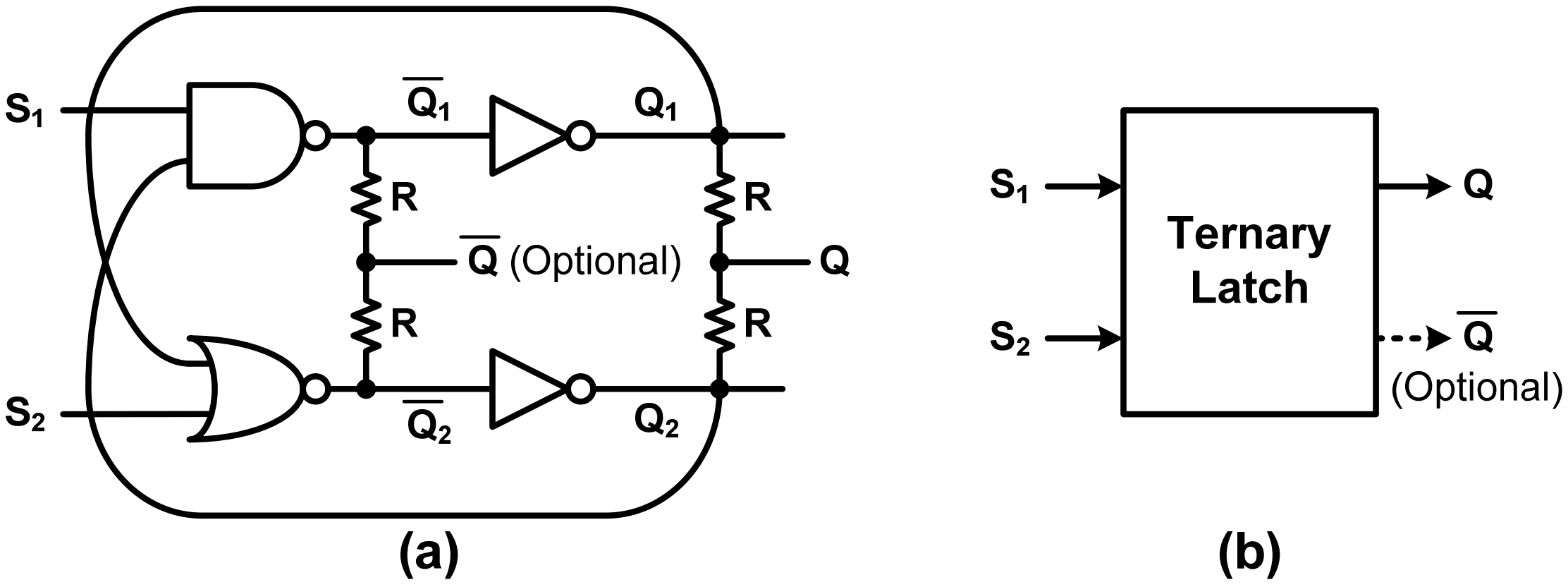

Fig. 2. The proposed ternary latch with the optional $\overline{Q}$ output, (a) Circuit, (b) Symbol

The ternary output, $Q$, is the average of the binary output values, $Q_1$ and $Q_2$ (Eq. 3). Table 1 shows that $Q_1$ and $Q_2$ become stable after 2Δt (Δt is the propagation delay of the logical gates), except when $S_1$ is *'High'* and $S_2$ is *'Low'*. This unstable situation is impractical. However, it is not actually even required for our purpose. It can easily be kept from happening. The other three combinations fulfill the requirements for constructing a ternary latch:

$$Q = \frac{Q_1 + Q_2}{2} \tag{3}$$

TABLE I
TIME-DEPENDENT TRUTH TABLE OF THE PROPOSED TERNARY LATCH

| $S_1$ | $S_2$ | $Q_1$ | $Q_2$ | $Q_1$+Δt | $Q_2$+Δt | $Q_1$+2Δt | $Q_2$+2Δt | $Q_1$+3Δt | $Q_2$+3Δt |
|---|---|---|---|---|---|---|---|---|---|
| Low | Low | 0 | 0 | 0 | 0 | 0 | 0 | 0 | 0 |
| Low | Low | 0 | $V_{DD}$ | 0 | 0 | 0 | 0 | 0 | 0 |
| Low | Low | $V_{DD}$ | 0 | 0 | $V_{DD}$ | 0 (Q = 0) | 0 | 0 | 0 |
| Low | Low | $V_{DD}$ | $V_{DD}$ | 0 | $V_{DD}$ | 0 | 0 | 0 | 0 |
| Low | High | 0 | 0 | 0 | $V_{DD}$ | 0 | $V_{DD}$ | 0 | $V_{DD}$ |
| Low | High | 0 | $V_{DD}$ | 0 | $V_{DD}$ | 0 | $V_{DD}$ | 0 | $V_{DD}$ |
| Low | High | $V_{DD}$ | 0 | 0 | $V_{DD}$ | 0 (Q = 1) | $V_{DD}$ | 0 | $V_{DD}$ |
| Low | High | $V_{DD}$ | $V_{DD}$ | 0 | $V_{DD}$ | 0 | $V_{DD}$ | 0 | $V_{DD}$ |
| High | Low | 0 | 0 | 0 | 0 | 0 | 0 | 0 | 0 |
| High | Low | 0 | $V_{DD}$ | $V_{DD}$ | 0 | 0 | $V_{DD}$ | $V_{DD}$ | 0 |
| High | Low | $V_{DD}$ | 0 | 0 | $V_{DD}$ | $V_{DD}$ | 0 | 0 | $V_{DD}$ |
| High | Low | $V_{DD}$ | $V_{DD}$ | $V_{DD}$ | $V_{DD}$ | $V_{DD}$ | $V_{DD}$ | $V_{DD}$ | $V_{DD}$ |
| High | High | 0 | 0 | 0 | $V_{DD}$ | $V_{DD}$ | $V_{DD}$ | $V_{DD}$ | $V_{DD}$ |
| High | High | 0 | $V_{DD}$ | $V_{DD}$ | $V_{DD}$ | $V_{DD}$ (Q = 2) | $V_{DD}$ | $V_{DD}$ | $V_{DD}$ |
| High | High | $V_{DD}$ | 0 | 0 | $V_{DD}$ | $V_{DD}$ | $V_{DD}$ | $V_{DD}$ | $V_{DD}$ |
| High | High | $V_{DD}$ | $V_{DD}$ | $V_{DD}$ | $V_{DD}$ | $V_{DD}$ | $V_{DD}$ | $V_{DD}$ | $V_{DD}$ |

- $S_1$ = *'Low'* and $S_2$ = *'Low'*: $Q_1 = Q_2 = 0 \rightarrow Q = \frac{0+0}{2} = 0$ (Logic '0')

- $S_1$ = *'Low'* and $S_2$ = *'High'*: $Q_1 = 0$, $Q_2 = \mathrm{V_{DD}} \rightarrow Q = \frac{0+V_{DD}}{2} = \frac{V_{DD}}{2}$ (Logic '1')

- $S_1$ = *'High'* and $S_2$ = *'High'*: $Q_1 = Q_2 = \mathrm{V_{DD}} \rightarrow Q = \frac{V_{DD}+V_{DD}}{2} = V_{DD}$ (Logic '2')

Ternary latches in [16] and [17] produce two complementary outputs at the same time. Their parallel production is the result of two inevitable voltage divisions, which occur within the ternary gates. The more voltage division occurs, the more static power dissipates. Note that the outputs, $Q$ and $\overline{Q}$, might not always simultaneously be required. The proposed design provides an opportunity for circuit designers to make a decision as to whether to produce both of them or not. This is an important option which is not available in the previous designs. This unique attribute is illustrated in figures by marking one the outputs with *'Optional'*.

### *2.2 The Proposed Ternary Level-Triggered Flip-Flap-Flop*

The proposed ternary latch can simply be converted to a Flip-Flap-Flop by involving the clock signal. Table 2 shows how the transformation process is possible. Note that the latch circuit does not have a *'Hold'* state, but it needs to hold whatever value is stored in the memory when the clock signal is *'Low'*. As soon as it becomes *'High'*, the input signals $Z_1$ and $Z_2$ determine the value of the latch by setting the appropriate values for $S_1$ and $S_2$.

Whenever clock is *'Low'*, $S_1$ and $S_2$ are set in the way that $Q$ remains unchanged regardless of the values of $Z_1$ and $Z_2$. As mentioned earlier, the unstable situation of $S_1$ = *'High'* and $S_0$ = *'Low'* must always be avoided. When clock becomes *'High'*, $Z_1$ and $Z_2$ determine the new value of $Q$. If $Z_1 \neq Z_2$, logic '1' will be stored in the memory. Equations 4 and 5 are first obtained from Table 2, and then rewritten in the CMOS logic style compatible form. Table 3 shows the truth table of the proposed ternary level-triggered F.F.F., whose circuit and block diagram are also depicted in Figs. 3a and 3b, respectively.

$$S_1 = \overline{CLK}.Q_1.Q_2 + CLK.Z_1.Z_2 = \overline{(CLK + \overline{Q_1} + \overline{Q_2})(\overline{CLK} + \overline{Z_1} + \overline{Z_2})} \tag{4}$$

$$S_2 = \overline{CLK}(Q_1 + Q_2) + CLK(Z_1 + Z_2) = \overline{(CLK + \overline{Q_1}.\overline{Q_2})(\overline{CLK} + \overline{Z_1}.\overline{Z_2})} \tag{5}$$

TABLE II
CONVERTING THE PROPOSED TERNARY LATCH TO A LEVEL-TRIGGERED F.F.F.

| CLK | Q | $Z_1$ | $Z_2$ | $S_1$ | $S_2$ |
|---|---|---|---|---|---|
| Low | 0 | × | × | Low | Low |
| Low | 1 | × | × | Low | High |
| Low | 2 | × | × | High | High |
| High | × | Low | Low | Low | Low |
| High | × | Low | High | Low | High |
| High | × | High | Low | Low | High |
| High | × | High | High | High | High |

TABLE III
TRUTH TABLE OF THE PROPOSED TERNARY LEVEL-TRIGGERED F.F.F.

| CLK | $Z_1$ | $Z_2$ | Q | State |
|---|---|---|---|---|
| High | Low | Low | 0 | Set '0' |
| High | Low | High | 1 | Set '1' |
| High | High | Low | 1 | Set '1' |
| High | High | High | 2 | Set '2' |

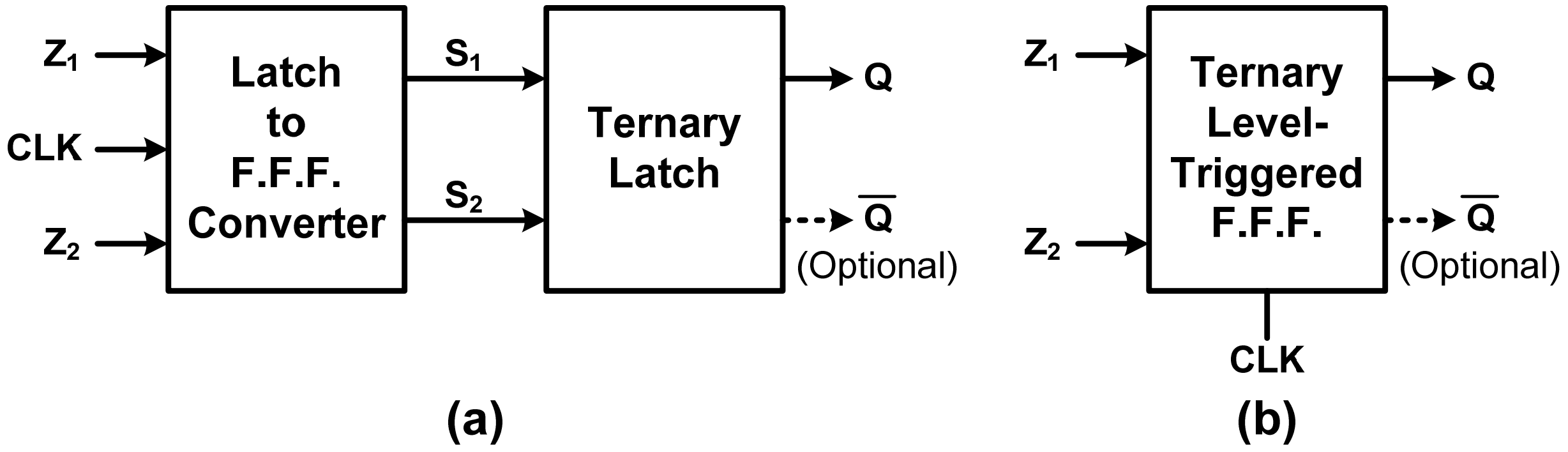

Fig. 3. The proposed ternary level-triggered Flip-Flap-Flop, (a) Circuit, (b) Symbol

*2.3 The Proposed Ternary Edge-Triggered Flip-Flap-Flop*

The level-triggered F.F.F. is convertible to an edge-triggered one by simply using the *master-slave* structure (Fig. 4a). The first(master) and second(slave) F.F.F. receive $\overline{CLK}$ and *CLK*, respectively. This allows the *master* to store the input value when the clock signal transitions from *'High'* to *'Low'*. At the same time, the output of the *slave* is locked. When the clock signal transitions vice versa, the signal captured by the now locked *master* passes through the *slave*.

Considering Eqs. 4 and 5, one can clearly understand that the *slave* requires $\overline{Z_1}$ and $\overline{Z_2}$ , not $Z_1$ and $Z_2$. Therefore, instead of $Q_1$ and $Q_2$, the inverted signals ( $\overline{Q_1}$ and $\overline{Q_2}$ ) are passed through by the *master* (Fig. 4a). These signals are generated inside the ternary latch (Fig. 2a), and hence no extra inverters are required. Another noticeable point is that the voltage division in the *master* is superfluous, since the *slave* takes binary inputs. Therefore, there is only one voltage division in the whole block. Finally, ($Q_{1M}$, $Q_{2M}$) never becomes (*'High'*, *'Low'*) as this unstable situation is always avoided. Therefore, ( $\overline{Q_{1M}}$ , $\overline{Q_{2M}}$ ) never becomes (*'Low'*, *'High'*). Thus, subsequently, the input combination of $\overline{Z_{1S}}$ = *'Low'* and $\overline{Z_{2S}}$ = *'High'* never occurs, and hence Eqs. 4 and 5 can further be simplified. Equations 6 and 7 show their simplified versions. The same thing will happen for the inputs of the *master* block ( $\overline{Z_{1M}}$ and $\overline{Z_{2M}}$ ). As a result, the *master* can follow the same equations as well (Eqs. 6 and 7). The block diagram of the proposed ternary edge-triggered Flip-Flap-Flop is depicted in Fig. 4b.

$$S_1 = \overline{CLK}.Q_1.Q_2 + CLK.Z_1 = \overline{(CLK + \overline{Q_1} + \overline{Q_2})(\overline{CLK} + \overline{Z_1})} \qquad (6)$$

$$S_2 = \overline{CLK}(Q_1 + Q_2) + CLK.Z_2 = \overline{(CLK + \overline{Q_1}.\overline{Q_2})(\overline{CLK} + \overline{Z_2})} \qquad (7)$$

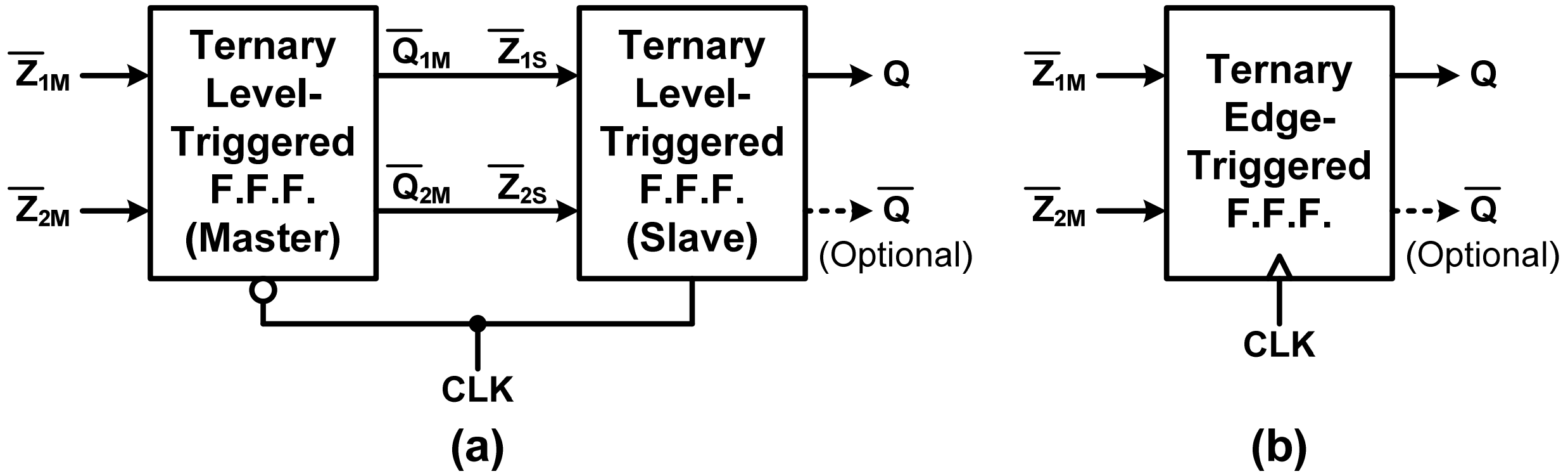

Fig. 4. The proposed ternary edge-triggered F.F.F. by using the *master-salve* structure, (a) Circuit, (b) Symbol

*2.4 The Proposed Ternary Edge-Triggered D Flip-Flap-Flop*

Eventually, the entire circuit is converted to a D F.F.F. with a single ternary input (Fig. 5a). A simple decoder, consisting of Positive and Negative Ternary Inverters (PTI and NTI), decodes the input signal and generates $\overline{D}+$ and $\overline{D}-$ to be connected to $\overline{Z}_{1M}$ and $\overline{Z}_{2M}$, respectively. Figure 5b exhibits the block diagram of the proposed ternary edge-triggered D F.F.F., whose truth table is also shown in Table 4. It is triggered by the rising edge of the clock, and stores the input value.

Figure 6 shows how the proposed design is implemented by 72 transistors. The entire implementation is in accord with the CMOS logic style, consisting of pull-up and pull-down networks. The PTI and NTI require transistors with a high threshold voltage (*High-$V_T$*) to be able to detect logic '1'. It is worth mentioning that multi-$V_t$ circuitry is an absolute necessity for MVL designs due to the fact that more than two voltage levels must be identified. However, except the decoder part, all of the transistors in the new design have a normal threshold. All of the transistors have the minimum feature size, except the p-type ones which are responsible for voltage division (Fig. 6). The width (W) of these transistors are increased to 90 nm (W = 90 nm) with the aim of dividing voltage precisely.

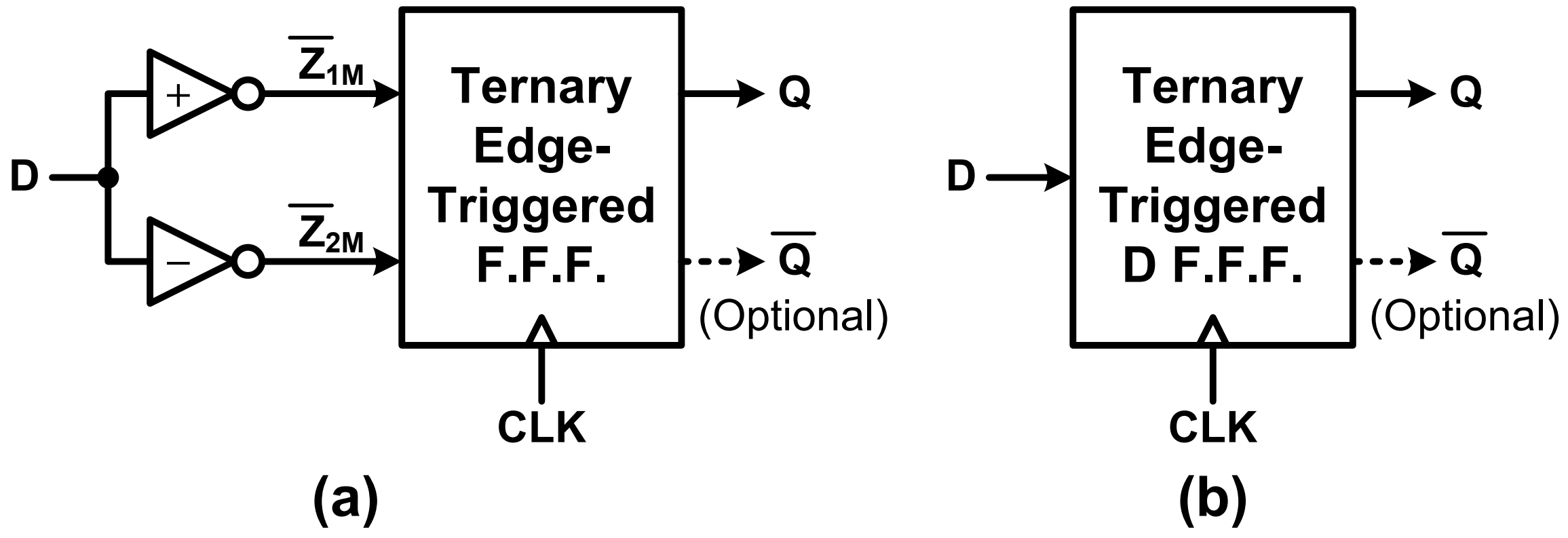

Fig. 5. The proposed ternary edge-triggered D Flip-Flap-Flop, (a) Circuit, (b) Symbol

**3. Simulation Results and Comparisons**

The proposed ternary edge-triggered D Flip-Flap-Flop and the ones presented in [16] and [17] are simulated by using HSPICE and 45 nm bulk-CMOS technology [18]. Simulations are carried out in 1 V power supply at room temperature. Table 5 shows simulation results and comparisons, which affirm the superiority of the proposed design, whose transient response is also shown in Fig. 7. The delay parameter is separately measured for all of the possible transitions (Table 5). In the worst-case transition ($0 \rightarrow 2$), the new structure operates approximately three and 2.6 times faster than [16] and [17], respectively.

The average power consumption is also measured during all of the transitions (Fig. 7). The proposed F.F.F. consumes the least power and it far surpasses other designs in terms of power consumption as well. Furthermore, in order to study power dissipation more accurately, static power dissipation is measured when the memory cell holds different logic values. The results are depicted in Table 5 as well. The highest amount of power dissipates when a ternary memory cell holds logic value '1', because constant voltage divisions are required to keep the inner and the output signals steady. Static current flows continuously from $V_{DD}$ to GND whenever voltage division occurs. The designs presented in [16] and [17] include 10 and four ternary components, respectively. Each one conducts static current individually. The more static current flows, the more static power dissipates. As it was mentioned before, the production of $Q/\overline{Q}$ is optional for the proposed design, and hence voltage division can occur only once. This is the reason why it consumes about 6.4 µW and 3.6 µW less static power than [16] and [17], respectively. Even if both outputs are generated, again power consumption is less than the other designs.

TABLE IV

TRUTH TABLE OF THE PROPOSED TERNARY EDGE-TRIGGERED D FLIP-FLAP-FLOP

| CLK | D | Q | State |
|---|---|---|---|
| ↑ | 0 | 0 | Set ‘0’ |
| ↑ | 1 | 1 | Set ‘1’ |
| ↑ | 2 | 2 | Set ‘2’ |

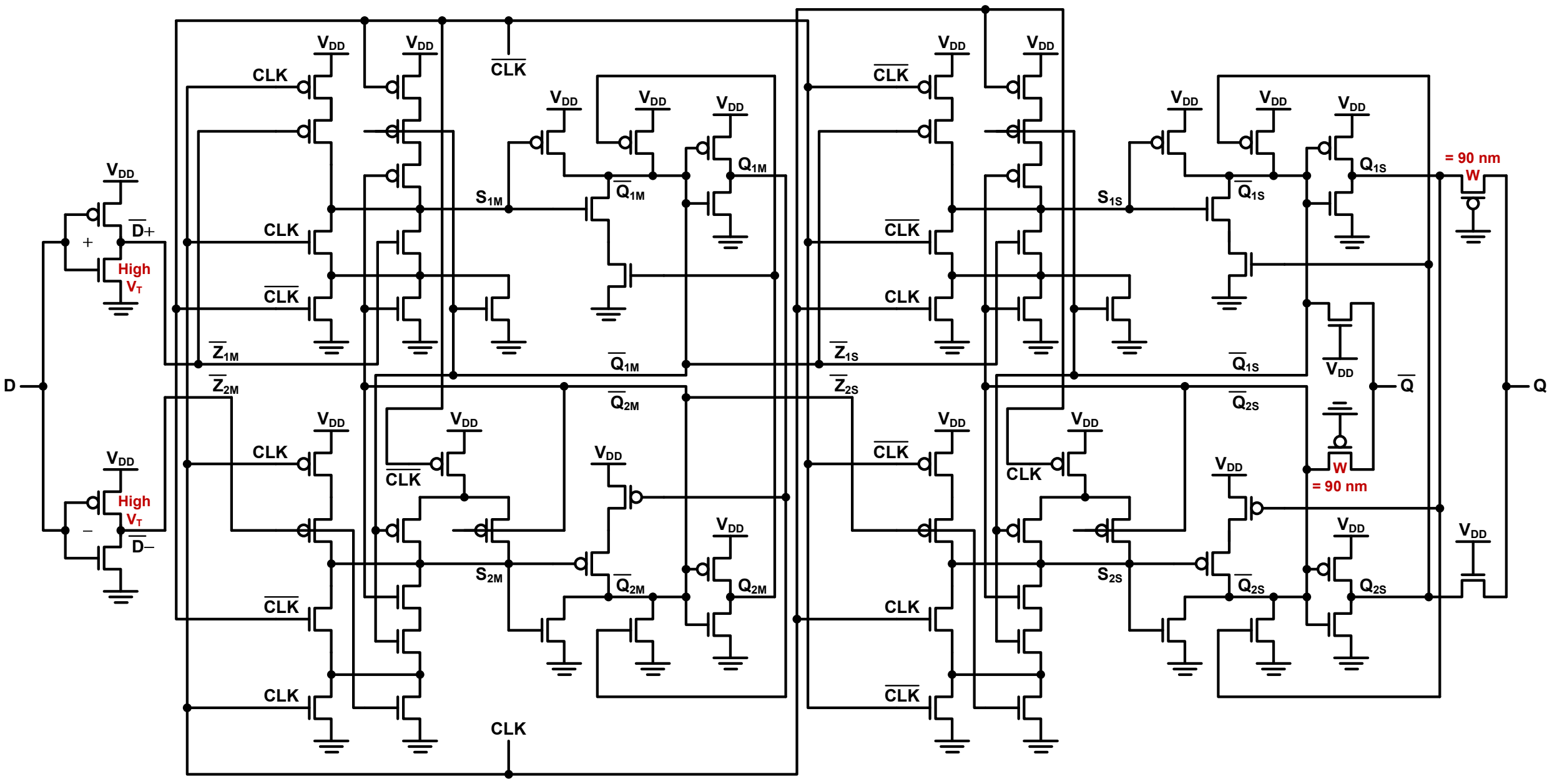

Fig. 6. Transistor-level implementation of the proposed ternary edge-triggered D F.F.F.

In digital electronics, the most important evaluating factor is energy consumption, which makes a balance between the delay and power parameters. It is calculated by Eq. 8. The proposed design has approximately 92% and 78% less energy consumption than [16] and [17], respectively. The proposed design is the most efficient ternary D F.F.F. although it does not have the fewest transistors. In spite of having the fewest transistors, the ternary D F.F.F. of [17] does not provide suitable performance in terms of other evaluating factors. Finally, the proposed design has 20 fewer transistors than [16].

$$Energy\ Consumption = Maximum\ Delay \times Average\ Power \tag{8}$$

TABLE V

SIMULATION RESULTS AND COMPARISONS

| Ternary Edge-Triggered D F.F.F. | | Proposed without $\overline{Q}$ | Proposed with $\overline{Q}$ | [16] | [17] |
|---|---|---|---|---|---|
| Delay (ps) | $0 \to 1$ | 63.772 | 69.196 | 90.504 | 99.416 |
| | $0 \to 2$ | 109.21 | 116.66 | 326.88 | 284.92 |
| | $1 \to 0$ | 57.982 | 61.687 | 463.01 | 116.62 |
| | $1 \to 2$ | 81.183 | 83.393 | 305.79 | 232.14 |
| | $2 \to 0$ | 86.445 | 95.044 | 489.30 | 24.641 |
| | $2 \to 1$ | 32.713 | 38.405 | 220.88 | 60.475 |
| Average Power (μW) | | 0.8306 | 1.1835 | 2.5143 | 1.4918 |
| Energy Consumption (aJ) | | 90.712 | 138.06 | 1230.2 | 425.04 |
| No. of Transistors | | 70 | 72 | 92 | 32 |
| Static Power (nW) | Holding ‘0’ | 63.232 | 62.969 | 216.61 | 17.406 |
| | Holding ‘1’ | 1249.8 | 2405.2 | 7667.9 | 4883.4 |
| | Holding ‘2’ | 98.712 | 101.98 | 206.35 | 16.792 |

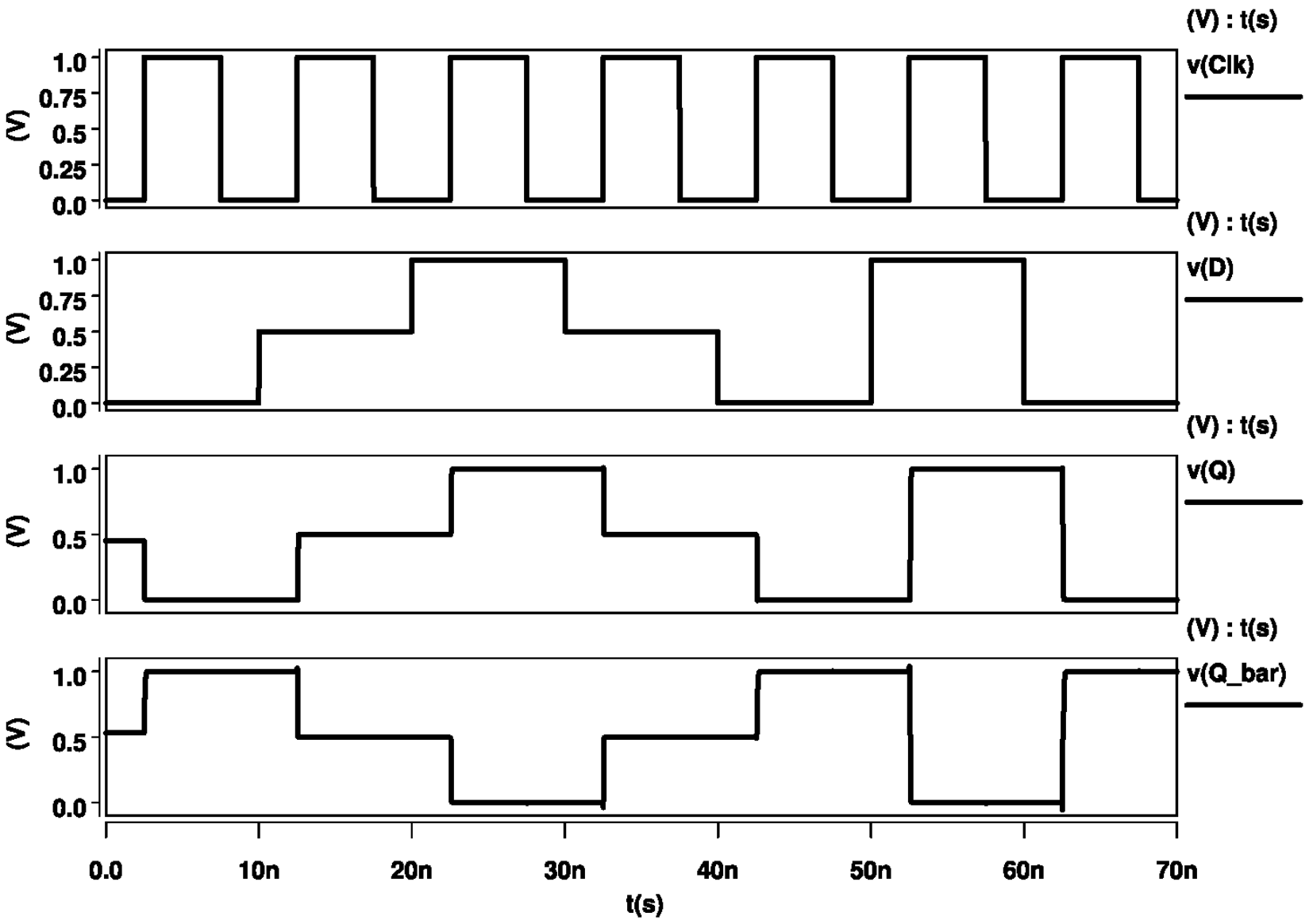

Fig. 7. The output waveforms of the proposed ternary edge-triggered D F.F.F.

## 4. Conclusion

A new ternary edge-triggered D Flip-Flap-Flop has been presented in this paper. The designing process has been explained in detail from the ternary latch to the final design. In comparison with the previous ternary D Flip-Flap-Flops, the new structure operates faster and consumes less power. The significant power reduction is mainly because of fewer voltage divisions, which happen inside the cell, especially when storing logic value '1'. The proposed design is mainly composed of binary circuits and it can efficiently be used in large ternary systems.

## References

[1] A. P. Dhande, V. T. Ingole, and V. R. Ghiye, "Fundamental concept of ternary logic," in *Ternary Digital Systems: Concepts and Applications*, SMGroup, 2014, pp. 1-17.

[2] R. C. Goncalves Da Silva, H. I. Boudinov, and L. Carro, "A low power high performance CMOS voltage-mode quaternary full adder," presented at IFIP Int. Conf. Very Large Scale Integration, Nice, Oct. 2006, pp. 187-191.

[3] K. C. Smith. (1981, Sept.). The prospects of multivalued logic: a technology and applications view. *IEEE Trans. Computer, C-30(9)*, pp. 619-627.

[4] P. C. Bella and A. Antoniou. (1984, Oct.). Low power dissipation MOS ternary logic family. *IEEE J. Solid-State Circuits, 19(5)*, pp. 739-749.

[5] E. Ozer, R. Sendag, and D. Gregg, "Multiple-valued logic buses for reducing bus energy in low-power systems," in *Proc. IEE Computers & Digital Techniques, 153(4)*, 2006, pp. 270-282.

[6] R. P. Hallworth and F. G. Heath, "Semiconductor circuits for ternary logic," in *Proc. IEE - Part C: Monographs, 109(15)*, 1962, pp. 219-225.

[7] E. Dubrova, "Multiple-valued logic in VLSI: challenges and opportunities," in *Proc. NORCHIP'99*, 1999, pp.340-350.

[8] J. Liang, L. Chen, J. Han, and F. Lombardi. (2014, Apr.). Design and evaluation of multiple valued logic gates using pseudo n-type carbon nanotube FETs. *IEEE Trans. Nanotechnology, 13(4)*, pp. 695-708.

[9] R. Faghih Mirzaee, T. Nikoubin, K. Navi, and O. Hashemipour. (2013, Dec.). Differential cascode voltage switch (DCVS) strategies by CNTFET technology for standard ternary logic. *Microelectronics J., 44 (12)*, pp. 1238-1250.

[10] S. Rezaie, R. Faghih Mirzaee, K. Navi, and O. Hashemipour. (2015, Dec.). From static ternary adders to high-performance race-free dynamic ones. *The J. Engineering*, pp. 1-12.

[11] B. Parhami. (2015, Mar.). Truncated ternary multipliers. *IET Computers & Digital Techniques, 9(2)*, pp. 101-105.
[12] K. You and K. Nepal. (2011, June). Design of a ternary static memory cell using carbon nanotube-based transistors. *IET Micro & Nano Letters, 6(6)*, pp. 381-385.
[13] S. Lin, Y. B. Kim, and F. Lombardi. (2012, Aug.). Design of ternary memory cell using CNTFETs. *IEEE Trans. Nanotechnology, 11(5)*, pp. 1019-1025.
[14] Ternary storage dynamic RAM, by T. Parks and D. D. Gaskins. (1995, July). *US Patent US5432735 A*.
[15] I. Arsovski, T. Chandler, and A. Sheikholeslami. (2003, Jan.). A ternary content-addressable memory (TCAM) based on 4T static storage and including a current-race sensing scheme. *IEEE J. Solid-State Circuits, 38(1)*, pp. 155-158.
[16] A. P. Dhande and V. T. Ingole. (2005, Apr.). Design of 3-valued R-S & D flip-flops based on simple ternary gates. *Int. J. Software Engineering and Knowledge Engineering, 15(2)*, pp. 411-417.
[17] M. H. Moaiyeri, M. Nasiri, and N. Khastoo. (2016, Mar.). An efficient ternary serial adder based on carbon nanotube FETs. *Engineering Science and Technology, an Int. J. 19(1)*, pp. 271-278.
[18] The Predictive Technology Model (PTM) website, Available at: http://ptm.asu.edu